\documentclass[twocolumn]{jetpl}

\usepackage{amssymb}
\usepackage{amsmath}
\usepackage{amsfonts}
\usepackage{graphicx}
\usepackage{bm}
\usepackage{color}
\usepackage{multirow}
\usepackage{cite}
\usepackage{ulem}

\begin{document}

\title{Comment on ``Noise in the helical edge channel anisotropically coupled to a local spin''}

\rtitle{Comment on ``Noise in the helical edge channel anisotropically coupled to a local spin''}

\sodtitle{Comment on ``Noise in the helical edge channel anisotropically coupled to a local spin''}

\author{I.\,S.\,Burmistrov$^{a}$\thanks{e-mail: burmi@itp.ac.ru.}, P.\,D.\,Kurilovich$^{b}$, V.\,D.\,Kurilovich$^{b}$}

%\dates{\today}{*}

\address{
$^{a}$ L. D. Landau Institute for Theoretical Physics RAS, 119334 Moscow, Russia \\
%$^{b}$ Raymond and Beverly Sackler School of Physics and Astronomy, Tel Aviv University, Tel Aviv 6997801, Israel \\
$^{b}$ Department of Physics, Yale University, New Haven, CT 06520, USA
}
\maketitle

In Ref.~\cite{NRS} the current noise in the helical edge channel anisotropically coupled to a local spin $1/2$ has been computed. In addition to the noise, a result for the backscattering current $I_{\rm bs}$ was reported. {The latter} formula (see Eq.~(7) of Ref.~\cite{NRS}) does not coincide with {the expression} for $I_{\rm bs}$ derived in our recent work (see Eq.~(22) of Ref.~\cite{KKBG}) for a general form of the exchange {interaction} matrix. Below we shall argue that, in general, the result of Ref.~\cite{NRS} for the backscattering current is \emph{erroneous}. Eq.~(7) {of Ref.~\cite{NRS}} gives the correct answer for {the} diagonal exchange matrix {only}. The incorrect result of Ref.~\cite{NRS} is {a} consequence 
of the assumption (which was also done in Ref.~\cite{Kimme2016}) that the density matrix of the impurity spin, $\rho_S$,  is diagonal in the eigenbasis of $S_z$ (see Eq.~(2) of Ref.~\cite{NRS}). As we demonstrated in Ref.~\cite{KKBG}, {a careful} analysis of the problem invalidates this assumption. 

In order to set notations, we define the Hamiltonian describing the exchange interaction between the helical edge states and a magnetic impurity as $H_{\rm int} = J_{jk} S_j s_k$, where 
$\bm{S}$ ($\bm{s}$) denotes the operator of the impurity spin ({the spin density of helical electrons}) and $J_{jk}$ is {a} $3\times 3$ exchange matrix. In Ref.~\cite{NRS} the following form of the exchange matrix was considered
\begin{equation}
J = \begin{pmatrix}
2(J_0+J_2) & 0 & 2J_a \\
0 & 2(J_0-J_2) & 0 \\
2J_1 & 0 & J_z
\end{pmatrix} .
\end{equation}
We note that in our paper \cite{KKBG} we used dimensionless exchange matrix $\mathcal{J}_{jk} = \nu J_{jk}$. Here $\nu = 1/(2\pi v)$ stands for the density of states {per edge mode and $v$ denotes the velocity of the helical states.} 

To illustrate our point we first consider the case  $J_2=J_1=0$ and the regime $V\gg T$. Then, according to Eq.~(7) of Ref.~\cite{NRS} the backscattering current is given {by} ($G_0 =e^2/h$)
\begin{equation}
I_{\rm bs}^{\rm NRS} =- G_0 T {J_a^2 }/{(2v^2)} .
\vspace{0.2cm}
\label{eq:eq1}
\end{equation}
This result should be contrasted with our result  \cite{KKBG}:
\begin{equation}
I_{\rm bs} = - G_0 \frac{V}{2v^2} \frac{2J_a^2J_0^2}{2J_a^2+J_z^2} .
\label{eq:eq2}
\end{equation}
In addition to a very different dependence of the backscattering current on {the} elements of the exchange matrix, Eq.~\eqref{eq:eq1} predicts saturation of the backscattering current at $V\gg T$ {whereas Eq.~\eqref{eq:eq2} does not}. This saturation occurs due to the full polarization of the magnetic impurity along $z$-axis by the applied voltage $V\gg T$. However, such a polarization is {a} consequence of {an} erroneous assumption that $\rho_S$ is diagonal in the eigenbasis of $S_z$. In fact, there are no physical reasons for the full polarization (along $z$-axis) to occur{:} the magnetic impurity remains partially polarized {in a direction tilted with respect to $z$-axis} for arbitrary large voltage (see discussion around Eq.~(26) in Ref.~\cite{KKBG}).

{To be more specific, the polarization along $z$-axis predicted by Ref.~\cite{NRS} follows from a claim that the dephasing of the impurity spin is mainly induced by the term $J_z S_z s_z$ in $H_{\rm int}$. However, the term $2J_a S_x s_z$ enters $H_{\rm int}$ on the equal grounds and thus has to be taken into consideration to properly account for the dephasing. In particular, if $J_z = 0$ the magnetic impurity gets polarized along $x$-axis for $V\gg T$. In this regime, the backscattering is induced by the term $2J_0 (S_x s_x + S_y s_y)$ in the Hamiltonian and is insensitive to the precise value of $J_a$. This is consistent with our Eq.~\eqref{eq:eq2} and not consistent with~Eq.~\eqref{eq:eq1}.}

Secondly, we consider the case $J_2=J_a=0$. Then, Eq.~(7) of  Ref.~\cite{NRS} predicts {a} linear in $V$ backscattering current
\begin{equation}
I_{\rm bs}^{\rm NRS} =-G_0  \frac{V}{4v^2} J_1^2 .
\label{eq:eq3} 
\end{equation}
Our result for this case coincides with Eq.~\eqref{eq:eq3} in the regime $V\gg T$. This occurs because the density matrix of the magnetic impurity $\rho_S$ is \emph{indeed diagonal} in the eigenbasis of $S_z$ for $J_a = 0$ and $V\gg T$. 

{In the regime of linear conductance ($V\ll \nu |J_{jk}| T$),} our result for the backscattering current reads
\begin{equation}
I_{\rm bs} =- G_0 \frac{V}{4v^2} \frac{J_1^2(J_z^2+2J_1^2)}{J_z^2+2J_1^2+4J_0^2} 
.
\label{eq:eq4}
\end{equation}
The discrepancy between Eqs.~\eqref{eq:eq3} and \eqref{eq:eq4} is due to  the non-diagonal structure of $\rho_S$ in the eigenbasis of $S_z$ {in the linear regime}. As one can see, our result \eqref{eq:eq4} transforms into Eq.~\eqref{eq:eq3} provided $|J_z|\gg |J_{0,1}|$, i.e., precisely {when} $\rho_S$ is diagonal in the eigenbasis of $S_z$.

To summarize, the result for the backscattering current reported in Ref.~\cite{NRS} is incorrect since its derivation  relies on the erroneous assumption. This also questions the result of Ref.~\cite{NRS} for the current noise (for the correct result for the shot noise {in the regime $V\gg T$} see Ref.~\cite{KKBGG}).

\end{document}